# Indo-US Research Collaboration: strengthening or declining?

**Jyoti Dua[1,2], Hiran H Lathabai[3], Vivek Kumar Singh[1,4,5]**

[1]Department of Computer Science, Banaras Hindu University, Varanasi-221005 (India)
[2]Department of Computer Science & Engineering, Pranveer Singh Institute of Technology, Kanpur-209305 (India)
[3]Amrita-CREATE, Amrita Vishwa Vidyapeetham, Amritapuri, Kerala-690525 (India)
[4] Department of Computer Science, University of Delhi, Delhi-110007 (India)
[5] Delhi School of Analytics, University of Delhi, Delhi-110007 (India)

{Emails: jyotidua1984@gmail.com, hiranhl007@gmail.com, vivek@cs.du.ac.in}

**Abstract:** Despite the importance of Indo-US research collaboration, it is intriguing to note that measurement and characterization of dynamics of Indo-US research collaboration is relatively underexplored. Therefore, in this work, we investigate major patterns in Indo-US collaboration with respect to certain key aspects using suitable scientometric notions and indicators. The research publication data for the last three decades (1990-2020) is obtained from Web of Science and analysed for the purpose. Results indicate an increase in absolute number of Indo-US collaborated papers over time, with an impressive share of about 1/3$^{rd}$ of India's total internationally collaborated research output. However, the proportionate share of Indo-US collaborated papers in India's internationally collaborated papers has declined over the time, as Indian researchers find new collaborating partners. Nevertheless, the collaboration with US is found to be highly rewarding in terms of citations and boost measures. Important insights and recommendations that may be helpful for shaping up new perspective on Indo-US collaboration policy are presented in this work.

## 1. Introduction

The importance of research collaboration is ever increasing due to its numerous advantages and mutual benefits it offers to the collaborating actors[16]. ICT revolution has significantly diluted many hinderances for collaborations including distance factor. Some studies witnessed the growth of international collaboration that has been increased linearly from last 2-3 decades[1, 2, 13, 15, 17]. Policymakers of different countries see collaboration as a valuable tool and designed various funding programs to foster the collaboration[6, 16, 21, 23]. Collaborative research played a crucial role for many developing countries to establish themselves at significant positions in global research landscape. For instance, India was placed on 5$^{th}$, 9$^{th}$ and 3$^{rd}$ position as per the studies carried out by Elsevier's Analytical Services[9], Clarivate Analytics[8] and the National Science Foundation (NSF) report on Science and Engineering indicators-2020[22]. This feat owes a lot to the perseverance devoted not only for the development of research ecosystem within the country but also for establishment and nurturing of international collaborations. Important studies related to international collaborations of India are discussed next.

India's overall international collaboration patterns in research has been analysed in several studies[3-5,14,19-20]. More recently, a study[11] measured and characterized international collaboration patterns in Indian scientific research, in a detailed study comprising research publication data for 20 years. Some studies also focused their attention on Indo-US collaboration, though with focus only on specific disciplines or thematic areas within disciplines like medicine and health[10], agriculture[18], etc. A previous work[7] analysed the overall scientific research trends between India and the US from the period 2007-2020 based on publications indexed in Scopus. The constant growth in collaboration was found from 2007 to 2016, with medicine and health sciences being the dominant area of collaboration. There are, however, many other aspects of Indo-US collaboration that are not yet explored. Given the importance of Indo-US collaboration in various respects, it is relevant to measure and characterize different aspects of the Indo-US collaboration, including looking at questions like whether it is strengthening or declining, and to what extent it has been beneficial for India.

Thus, this work intends to provide quantitative analysis of bilateral collaboration between India and the US for the last 31 years (1990-2020). The analysis provides the present status of Indo-US collaboration and how impactful the collaboration evolved, including relative intensity of collaboration. More precisely, the paper intends to answer the following research questions:

**RQ1:** What proportion of Indian international research collaboration is with US and whether it has increased or decreased with time?

**RQ2:** How has the relative intensity of collaboration of India and the US changed over time?

**RQ3:** How impactful has been the Indo-US research collaboration?

**RQ4:** In how many and which fields and subject areas do India and the US collaborate more?

## 2. Data & Method

This study analyses publications of India for the last three decades to address the research questions. Research publication data was downloaded from Web of Science (one of the most reliable and widely used scholarly databases) for the period 1990-2020 (31 years) using following search query:

*CU=India and PY=(1990-2020) and LA=English and (DT=Article or DT=Review)*

Where, CU field refers to 'country name', PY field to 'publication year', LA to the 'language' and DT to the 'Document Type'. The reason behind restricting publication data to 'Article" and "Review" is that they are the most representative document types reporting research contribution. During pre-processing, the duplicate values were removed based on DI (digital object identifier). This resulted in a final set of 863,204 unique publication records.

## 2.1 Dynamics of Proportion of Indo-US collaboration (RQ1)

There were 60 metadata fields in the publication record data downloaded, of which our analysis involved processing of information in DI, C1, WC and Z9 fields. C1 denotes author's affiliation, including affiliation country, and was used to categorize publication records into three groups: Indigenous, Indo-US collaborated papers, and internationally collaborated papers. Indigenous publications involved authors only from India. Those publications that had at least one author from the US and India are called Indo-US collaborated papers. In addition to Indian author, those publications that had at least one author from any other country are categorized as internationally collaborated papers (ICP).

A total of 648,475 publication records (out of total 863,204) were identified as indigenous papers, and 214,729 are internationally collaborated papers (ICP). Out of the ICP instances, a total of 69,243 papers are Indo-US research collaborated papers. The proportionate share of the different category of papers was analysed and recorded year-wise. The year-wise ICP and those involving Indo-US collaborative publications were analysed. Compound Annual Growth Rate (CAGR) of all the three categories of publications were also computed.

## 2.2 Dynamics of relative intensity of Indo-US collaboration (RQ2)

Next, the relative intensity of collaboration (RIC)[12] of India and the US with major collaborating countries was computed for the given period. For this purpose, the number of publication records for the 25 countries was retrieved from Web of Science. The RIC is defined as the ratio of the share of the collaborations of actors X and Y within all collaborations of X to the share of collaborations of Y within all collaborations of the system excluding collaborations of X. It is expressed as-

$$RIC(X,Y) = \frac{C_{xy} \times (T - C_X)}{C_x \times (C_y - C_{xy})} \quad (1)$$

where, $Cxy$ denotes the number of collaborations between two countries X and Y, $Cx$ is the total number of collaborations of country X, $Cy$ is the total number of collaborations of country Y and T represents the total number of pairwise collaborated publications of countries under study.

## 2.3 Overall impact of Indo-US collaboration in India's scholarly output (RQ3)

The effectiveness of collaboration was analysed in terms of boost in productivity and citation for India. Boost indicators and framework for determination of overall boost in productivity and impact of an actor (like country) due to international collaborations was introduced by[11]. In this work, with a slight modification of some of the indicators in the framework, the overall boost on productivity and impact of Indian scholarly output due to Indo-US collaborations can be found out. Relevant definitions of indicators are discussed next.

**Productivity boost ($\beta_P'$) of a country due to a bilateral-research collaboration:** It is the ratio of sum total of indigenous publications of a country and bilaterally collaborated publications with another country (in this case, Indo-US collaborated publications, making $T_{IP} + T_{IPUS}$) to the total indigenous productivity ($T_{IP}$) of a country (in this case, India), expressed in percentage.

$$\beta_P' = \left[\frac{(T_{IP} + T_{IPUS})}{TIP} - 1\right] \times 100 \% \quad (2)$$

The higher the value of $\beta_P$, the greater the boost productivity of a country due to its bilateral collaborating partner. The $\beta_P$ also indicates the extent on which a country is dependent on its collaborating partner. It is difficult to determine the ideal value of $\beta_P'$, but as per thumb rule, if $\beta_P' > 20\%$, there is a greater probability of over dependence considering that there might be multiple partners for a country, but more dependence is towards a particular country. If $\beta_P > 100\%$, this indicates that the country is on a high dependence on another country.

**Citation boost ($\beta_c'$) of a country due to a bilateral-research collaboration:** It is the ratio of sum total of citations together received by indigenous publications of a country and bilaterally collaborated publications with another country (in this case, Indo-US collaborated publications, making $T_{IC} + T_{ICUS}$) to the citations received by total indigenous publications ($T_{IC}$) of a country (in this case, India), expressed in percentage.

$$\beta_c' = \left[\frac{(T_{IC}+T_{ICUS})}{T_{IC}} - 1\right] \times 100\% \quad (3)$$

The rationale behind this is that the higher the value of $\beta_c'$, the greater the citation boost of a country due to its collaborating partner. It is difficult to determine the ideal value of $\beta_c'$, but as a rule of thumb, if $\beta_c' > 30\%$, there is a greater probability of over dependence for impact. If $\beta_c' > 100\%$, this indicates that the country is on a high dependence on another country for impactful scholarly production.

**Boost ratio of impact per unit boost in productivity ($\gamma_c'$):** It is the net boost of impact per unit boost of productivity due to bilateral collaborations.

$$\gamma_c' = \frac{\beta_{C'}}{\beta_{P'}} \quad (4)$$

If $\gamma_c' < 1$, collaborations are less rewarding and if $\gamma_c' > 1$, collaborations are rewarding. Greater the value of $\gamma_c'$, greater the benefit of collaboration. However, as high $\beta_C'$ can indicate higher dependency or prevalence of low impact of indigenous productivity.

**Citedness boost ($\beta_{rc}'$):** It is the ratio of sum total of citedness of indigenous and bilateral publications taken together to the citedness ratio of the indigenous publications. In Indo-US case, the bilateral citedness is represented as $r_{TIUS}$, making the expression of $\beta_{rc}'$ as:

$$\beta_{rc}' = \left[\frac{r_{TIUS}+r_{TI}}{r_{TI}} - 1\right] \times 100\% \quad (5)$$

where

$$r_{TIUS}' = \frac{total\ number\ of\ cited\ Indo-US\ publications}{total\ number\ of\ Indo-US\ publications} = \frac{T_{IPUS}\ cited}{T_{IPUS}} \quad (6)$$

&

$$r_{TI} = \frac{total\ number\ of\ cited\ indigenous\ publications}{total\ number\ of\ indigenous\ publications} = \frac{T_{IP}\ cited}{T_{IP}} \quad (7)$$

Citedness boost value greater than but close to 1 indicate that indigenous publications are also having considerably good citedness or capable of attracting citations. $\beta_{rc}'$ and $\beta_c'$ can be together used to determine whether a country's indigenous works are making enough impact.

**Boost ratio of impact per unit boost in citedness ($\delta_c'$):** It is the net boost of impact per unit boost of citedness due to bilateral collaborations.

$$\delta_c' = \frac{\beta_{C'}}{\beta_{rc'}} \quad (8)$$

The rationale behind this is that higher the value of $\delta_c'$, higher the effectiveness of collaboration. If the value of $\delta_c'$ is very high with $\beta_{rc}' < 1\%$, indicates that a majority of collaborations are of good quality and rewarding. In contrast, higher the value of $\delta_c'$ with $\beta_{rc}' > 1\%$, indicate less rewarding collaborations, therefore, these can be reviewed and decisions on whether to strengthen such collaborations or to minimize focus on such collaborations can be taken.

**2.4 Research area-wise growth pattern of Indo-US collaboration (RQ4)**

The Web of Science (WoS) categories of collaborated publications were analysed to see in which subject areas India and the US collaborated most. For this WC field in the metadata were analysed. Then, the total number of categories (out of 252 WoS categories) in which the two countries collaborated were identified and plotted year-wise. Secondly, the total number of papers in each WoS category was obtained and sorted in descending order to analyse the major subject areas with high collaboration volumes. Further, the top ten WoS categories were selected and listed.

**3. Results**

*RQ1: What proportion of Indian international research collaboration is with US and whether it has increased or decreased with time?*

For answering this question, the proportion of indigenous, ICP, and Indo-US instances of Indian research output are identified in a year-wise manner from 1990 to 2020. The observation is collected in **Table 1**, which presents volume and percentage of the total research output, indigenous papers, ICP, and Indo-US share in ICP during the last 31 years.

It can be observed that in the year 1990, a total of 348 papers out of a total of 2,963 papers published by India included international collaboration, which constituted 11.74% of the total number of papers. Out of 348 internationally collaborated papers, there were 156 Indo-US shares that were observed in 1990. This constitutes 44.83% of the total ICP. By the year 2020, as the number of internationally collaborated papers has increased to 26,295 (which is 32.08% of the total research output), the share of Indo-US collaboration has increased in volume but its percentage has decreased to 27.77%.

In the overall 31 years, an increase in Indo-US collaboration volume can be observed, but at the same time, a decrease in its proportionate share can also be seen. It can be observed that the Indo-US share at 2020 in ICP is only 32.25%. In the year 1990, the collaboration percentage was close to 45% but dropped in 1991 to 40%. In the year 1993, the percentage of share increases again and reaches close to the percentage similar to 1990. We can observe that from 1990 to 1999, the share was within 40%, but later, that is, from 2000 to 2020, the share eventually dropped below 40% and declined continuously. These patterns indicate that India has found newer collaborating partners in this period.

**Table 1: Total papers, internationally collaborated papers and Indo-US collaboration share for India (1990-2020)**

| Year | Total papers | Indigenous | | Inter-Collaborated | | Indo-US Share in ICP | |
|---|---|---|---|---|---|---|---|
| | | Number of papers | Percentage | Number of papers | Percentage | Number of papers | Percentage* |
| 1990 | 2963 | 2615 | 88.26 | 348 | 11.74 | 156 | 44.83 |
| 1991 | 3466 | 3009 | 86.81 | 457 | 13.19 | 188 | 41.14 |
| 1992 | 3891 | 3369 | 86.58 | 522 | 13.42 | 221 | 42.34 |
| 1993 | 4028 | 3468 | 86.10 | 560 | 13.90 | 253 | 45.18 |
| 1994 | 4246 | 3630 | 85.49 | 616 | 14.51 | 271 | 43.99 |
| 1995 | 4473 | 3785 | 84.62 | 688 | 15.38 | 278 | 40.41 |
| 1996 | 5335 | 4492 | 84.20 | 843 | 15.80 | 353 | 41.87 |
| 1997 | 6137 | 5193 | 84.62 | 944 | 15.38 | 404 | 42.80 |
| 1998 | 9960 | 8170 | 82.03 | 1790 | 17.97 | 735 | 41.06 |
| 1999 | 10417 | 8569 | 82.26 | 1848 | 17.74 | 700 | 37.88 |
| 2000 | 10579 | 8508 | 80.42 | 2071 | 19.58 | 788 | 38.05 |
| 2001 | 11357 | 9003 | 79.27 | 2354 | 20.73 | 889 | 37.77 |
| 2002 | 13023 | 10317 | 79.22 | 2706 | 20.78 | 960 | 35.48 |
| 2003 | 14237 | 11251 | 79.03 | 2986 | 20.97 | 1040 | 34.83 |
| 2004 | 15816 | 12407 | 78.45 | 3409 | 21.55 | 1243 | 36.46 |
| 2005 | 17945 | 14107 | 78.61 | 3838 | 21.39 | 1336 | 34.81 |
| 2006 | 21251 | 16628 | 78.25 | 4623 | 21.75 | 1573 | 34.03 |
| 2007 | 24433 | 19247 | 78.77 | 5186 | 21.23 | 1730 | 33.36 |
| 2008 | 27646 | 21747 | 78.66 | 5899 | 21.34 | 2002 | 33.94 |
| 2009 | 25573 | 19814 | 77.48 | 5759 | 22.52 | 1879 | 32.63 |
| 2010 | 34508 | 26419 | 76.56 | 8089 | 23.44 | 2669 | 33.00 |
| 2011 | 38222 | 29151 | 76.27 | 9071 | 23.73 | 2941 | 32.42 |
| 2012 | 42330 | 32310 | 76.33 | 10020 | 23.67 | 3370 | 33.63 |
| 2013 | 47302 | 36116 | 76.35 | 11186 | 23.65 | 3649 | 32.62 |
| 2014 | 53349 | 40542 | 75.99 | 12807 | 24.01 | 4171 | 32.57 |
| 2015 | 55542 | 41771 | 75.21 | 13771 | 24.79 | 4396 | 31.92 |
| 2016 | 61477 | 45065 | 73.30 | 16412 | 26.70 | 5339 | 32.53 |
| 2017 | 66016 | 48270 | 73.12 | 17746 | 26.88 | 5595 | 31.53 |
| 2018 | 70767 | 50897 | 71.92 | 19870 | 28.08 | 6254 | 31.47 |
| 2019 | 74949 | 52934 | 70.63 | 22015 | 29.37 | 6558 | 29.79 |
| 2020 | 81966 | 55671 | 67.92 | 26295 | 32.08 | 7302 | 27.77 |
| Total | 863204 | 648475 | 75.12 | 214729 | 24.88 | 69243 | 32.25 |
| CAGR | 11.70% | 10.73% | | 15.51% | | 13.68% | |

*Indo-US_Collaborated percentage=No. of Indo-US papers /No. of India's Inter-collaborated papers

*RQ2: How has the relative intensity of collaboration of India and the US changed over time?*

In order to analyse the collaborating patterns of the Indo-US collaboration network, the relative intensity of collaboration (RIC) of India and the US is computed with respect to their major collaborating partners. India's major collaborators arranged in descending order are- the US, Germany, England, South Korea, China, France, Japan, Australia, Italy, Canada. The RIC plots can now be interpreted for these major countries. The dynamics of RIC indicator for India with respect to its major collaborating partners is plotted in **Figure 1a**.

It can be identified that the RIC with respect to the US, Japan and China almost shows a similar upward and downward trend up to 1999. But, later during the period 2000-2020, many different patterns can be identified with these countries. RIC with respect to US declined. In case of Japan, mixed trend is found, whereas RIC with Australia shows upward trend (at low rate) since 2010. After abrupt decline since 1999, RIC with respect to China is found to rise since

2016. RIC with respect to England shows a consistent rise since 2010 but at very low rate. Germany, France and Canada show a decline. RIC with Italy was increasing since 2010, but slightly declined after 2018. RIC increase with respect to South Korea witnessed maximum rise, though slight dip is there since 2018. Apart from these, some other top collaborative countries with India (that are not shown in Fig 1a) also offer some promising growth in the RICs. Upward trend with respect to Saudi Arabia, Israel, and Taiwan can be identified, whereas Malaysia, South Africa, Brazil, the Netherlands, Sweden, Switzerland, Belgium, and Denmark show decline.

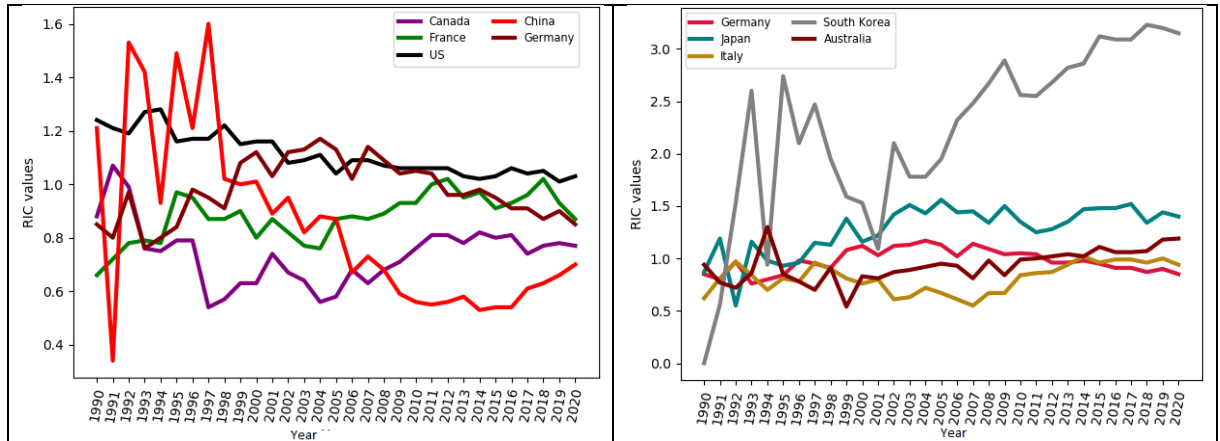

**Figure 1a: RIC of India with its major collaborating partners**

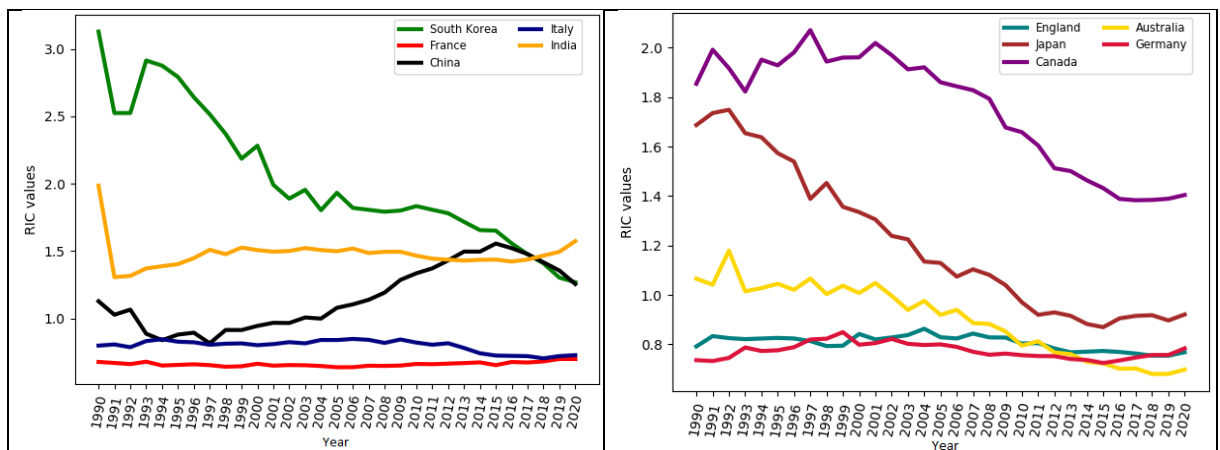

**Figure 1b: RIC of the US with its major collaborating partners**

US's major collaborators arranged in descending order are- China, Germany, England, India, Canada, France, Japan, Italy, Australia, South Korea. The RIC plots can now be interpreted for these major countries. The RIC plot for the US with respect to its major collaborating partners is plotted in **Figure 1b**.

It can be identified that the RIC with respect to India and Germany shows an upward trend post 2015, whereas RIC with South Korea registered the highest overall decline. RIC with Canada and Japan also has shown similar pattern, got slightly consistent rise from 2018 to 2020. Australia, and China show a downward trend in RIC, whereas, after showing a mixed trend, Germany, England and Italy show an upward trend at 2020. With France, RIC had been consistent till 2015 and afterwards slight upward growth is registered. Apart from these, other prominent collaborative countries with respect to US (not shown in fig 1b) are also discussed. Russia, Malaysia, Taiwan, Saudi Arabia and Poland show a decline whereas Israel, Spain, France, Brazil, South Africa and Belgium show growth in RIC. Countries like Switzerland, Denmark and Netherlands have shown mixed trend till 2015 and afterwards grown slightly.

Analysis of RICs of both countries (India and the US) revealed that both countries are improving and maintaining collaborative ties with some nations, reclaiming strength of ties with some others after decline and consistently loosening ties with some other. As the partners where RIC is improving (and also declining) is spread across different continents, it is difficult to attribute any geopolitical influence on it. In the case of Indo-US bilateral research relationship, the most intriguing fact is that India's RIC with US is declining while, US's RIC with India is gradually rising.

*RQ3: How impactful has been the Indo-US research collaboration?*

The next question analysed is: *How impactful is the Indo-US collaborated research in an Indian publication?* To answer this, cited percentage and citations per paper (CPP) for indigenous, inter-collaborated and Indo-US categories of research papers is computed. It can be observed in **Table 2** that Indo-US participated papers received slightly more cited% (88.93%) than ICP, which received 87.47%. Thus, it is found that there is very slight difference of potential impact between ICP and Indo-US research outputs. Similarly, for citations per paper it was observed that the value for ICP is 22.32 citations per paper as compared to 31.44 citations per paper for Indo-US collaborated papers. The citation per paper for Non-ICP or indigenous papers is lower than ICP and Indo-US. The citation per paper for total research output (TP) is 16.78. Thus, Indo-US collaboration gets a clear advantage in terms of citation impact over ICP and indigenous impact.

**Table 2: Citation Impact of Indo-US collaboration**

|  | Category of papers | | | |
| --- | --- | --- | --- | --- |
|  | TP | Non-ICP | ICP | Indo-US Collaborated |
| No. of papers | 863204 | 648475 | 214729 | 69243 |
| Cited % | 86.19 | 85.76 | 87.47 | 88.93 |
| Citations per paper | 16.78 | 14.95 | 22.32 | 31.44 |

**Boost in Productivity and Impact**

**Productivity boost of India**

$$\beta_P' = \left[\frac{717718}{648475} - 1\right] \times 100\ \% = 10.7\ \%$$

As we can observe India's productivity boost is less than 20%, India is not too much reliant on the US collaborations for productivity.

**Citations boost for India**,

$$\beta_C' = \left[\frac{11869018}{9692229} - 1\right] \times 100\ \% = 22.5\ \%$$

This value indicates that India is not too much reliant on the US collaborations for citations. Though the scholarly system of India is productive, it slightly fails somewhere to receive necessary impact when collaborated with the US. Further, boost ratio of impact was computed to find rewarding collaborations on unit % increase in productivity.

$$\gamma_c{'} = \frac{\beta_c{'}}{\beta_P{'}} = \frac{22.5}{10.7} = \mathbf{2.1}$$

Thus, for India, for each 1 % boost of productivity achieved through collaboration, 2.1 % boost in citations is achieved. This indicates collaboration of India with the US is highly rewarding.

For further clarity, the boost in citedness is achieved with the US collaborations is also analysed.

Citedness boost ($\boldsymbol{\beta_{rc}{'}}$) of India can be computed in the following way:

$$r_{TIUS}{'} = \frac{617727}{717718} = \mathbf{0.861}$$

$$r_{TI}{'} = \frac{556149}{648475} = \mathbf{0.8576}$$

$$\boldsymbol{\beta_{rc}{'}} = \left[\frac{r_{TIUS}{'}}{r_{TI}{'}} - 1\right] \times 100\ \% = \left[\frac{0.861}{0.8576} - 1\right] \times 100\ \% = \mathbf{0.36\ \%}$$

Citedness boost ($\boldsymbol{\beta_{rc}{'}}$) computed is found to be less than 1%, hinting the indigenous scholarly works are capable of attracting citations, but to know the difference in capability of Indo-US collaborated works to attract citations in huge volume, boost ratio of impact per unit boost in citedness needs to be computed.

$$\delta_c = \frac{\beta_c{'}}{\beta_{rc}{'}} = \frac{22.5}{0.36} = \mathbf{63.03}$$

In the case of India, per unit percentage boost in citedness achieved by the US collaboration, citations can improve by almost 63%. Thus, Indian Indigenous scholarly ecosystem is not weak in citedness but failing to attract citations at a deserved level. Indian scholars should act more diligently and adopt effective strategies to maximize the reach of their research work after maintaining top-notch quality and choosing impactful journals for dissemination.

*RQ4: In how many and which fields and subject areas do India and the US collaborate more?*

**Figure 2** presents the number of fields in which India and the US collaborated. It has been observed that the number of collaborative fields has increased during the period 1990-2020. The distribution of co-authored papers in various Web of Science (WoS) subject categories is slightly uneven. In the year 1990, the number of categories in which India and the US collaborated was below 100. Further, by 2020, the two countries collaborated in more than 200 fields. For instance, Astronomy & Astrophysics; Biochemistry & Molecular Biology; Biology Chemistry, Physical, and Chemistry, Multidisciplinary etc., are few fields in which India and the US collaborated for three decades. Agricultural Engineering, Agriculture, Dairy & Animal Science, Allergy, Criminology & Penology, Cell & Tissue Engineering, Dentistry, Oral Surgery, and Medicine, etc., are few fields in which India and the US have not initially collaborated, but by the end of the 1990s the two countries started collaborating. Classics, Dance, Folklore, Literary Reviews, Literary Theory & Criticism, Literature, American, Literature, British Literature, Romance, Mediaeval & Renaissance Studies, Poetry, and Psychology, Psychoanalysis are some fields of WoS in which India and the US never collaborated. While looking at these patterns, it may however be also kept in mind that number of WoS categories have changed slightly during the period of 30 years.

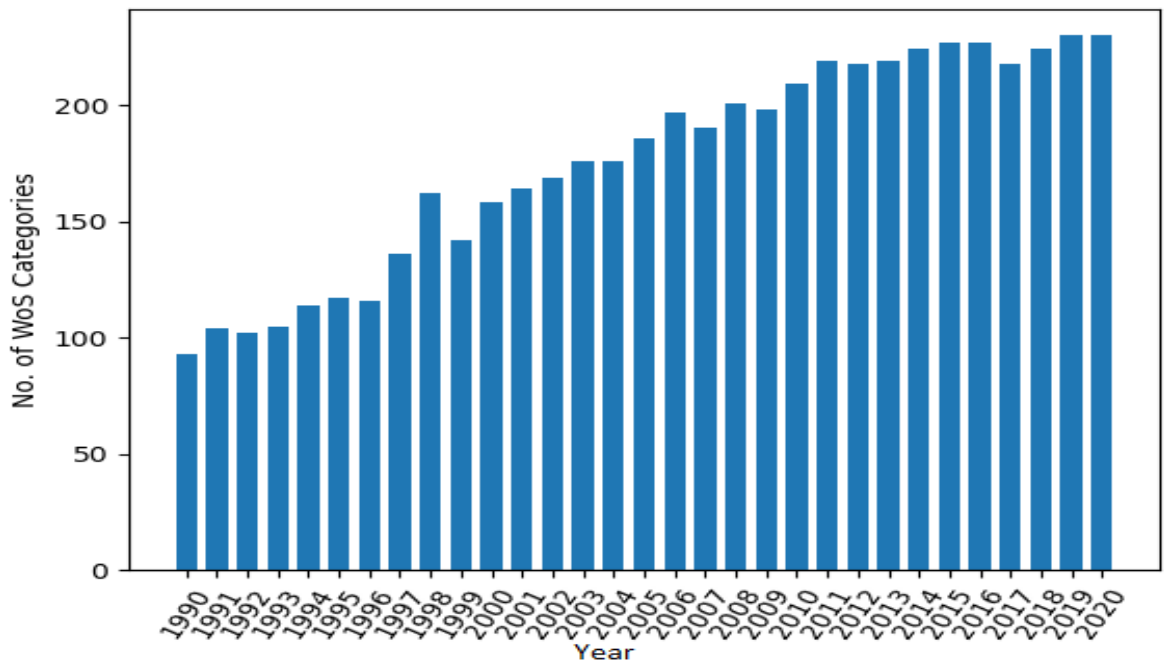

**Figure 2: Number of WoS categories in which India and the US collaborated**

**Table 3: WoS categories with high Indo-US research collaboration volume**

| WoS Categories | No. of Papers |
|---|---|
| Astronomy & Astrophysics | 5617 |
| Biology | 5495 |
| Materials Science, Multidisciplinary | 4496 |
| Physics, Particles & Fields | 3991 |
| Biochemistry & Molecular Biology | 3468 |
| Physics, Applied | 2953 |
| Chemistry, Physical | 2806 |
| Multidisciplinary Sciences | 2744 |
| Microbiology | 2549 |
| Chemistry, Multidisciplinary | 2520 |

Finally, **Table 3** presents the top 10 subject areas (arranged in descending order of collaborated papers). It is seen that Astronomy & Astrophysics and Biology subject areas have relatively more Indo-US collaborative papers, with 5,617 and 5,495 respectively. The other top collaborative thematic categories are Material Science, Multidisciplinary, Physics, Particles &

Fields, and Biochemistry & Molecular Biology, etc. Thus, Indo-US collaborations are mostly strong in impactful scientific disciplines barring a few multidisciplinary ones.

## 4. Discussion

The article provides a quantitative analysis of Indo-US research collaboration patterns during the last three decades (1990–2020). Profound emphasis (through RQs 1 to 4) was laid on determination of (i) dynamics of proportion of Indo-US collaboration in India's international collaboration, (ii) dynamics of relative intensity of collaboration of India as well as US, (iii) boost in productivity and impact for India due to collaboration with US and (iv) major areas /subject categories of mutual interest that flourished under Indo-US collaboration.

From our analyses, it is found that Indo-US collaborations is improving but, in a rate, lesser than that of overall international collaborations of India (from investigation of RQ1). This can be a hint of Indian actor's preference to other collaboration partners. Further evidence to this is obtained while addressing RQ2, as relative intensity of collaboration with the US is found to be declining whereas India's RIC with some other partners is increasing. However, the US's RIC with India is steadily increasing after a dip in early 1990's, indicates they value Indian cooperation highly. Analysis of institutional overall boost (determined as per RQ3) achieved in India's productivity and impact due to collaboration with US indicates that indigenous scholarly ecosystem is sufficiently strong, but not as impactful as it should have been. To address the impact limitation of indigenous research, while collaborating with highly rewarding partners like the US, Indian actors should also focus on learning how to plan, do, determine most suitable outlet, effectively communicate and publish, and effective use of visibility maximization.

As collaboration with the US is highly rewarding for India, it should focus on strengthening ties with the US and explore new subject areas or categories other than the currently emphasized areas (determined as per RQ4), as the US highly values India as a research partner. Major hindering factors for this should be identified at different levels and addressed. Our own rudimentary analysis of collaboration pattern at individual and institutional level during the period can act as starting point in this direction. We also tried to analyse the dynamics of publications with Indian authors as first-author. It revealed no particular pattern or trend, from 39.1% in 1990, improved to 45.02% in 1994 and progressed slight dips and ups and reached at 41.81% in 2020. This indicates the need for more proactive effort at individual level to reach out for, to maintain and strengthen collaborations with the US. When it comes to institutional collaboration, during decades 1990-99, 2000-2010 and 2011-2020, Tata Institute of Fundamental Research (TIFR), Delhi University (DU) and Punjab University (PU) are found to occupy top three positions, but they switched positions among themselves over time. Also, their collaborating partner preference changes are evident as ties between new partners got strengthened over the older ones. For instance, if University of Michigan, University of California, University of Hawaii, etc, were major partners at 1990-99, Princeton University, University of Illinois, Northeastern University, etc., dominated the Indo-US partnership landscape of the top Indian institutions. This further changed in 2011-2020, where Ohio State University, Wayne State University, University of Mississippi, etc., are found to be major partners. Apart from these, institutions like IISc, Bhabha Atomic Research Centre (BARC) were major institutions with high Indo-US collaborations in 1990-99 and 2000-2009. However, during 2010-2020, other new institutions like Homi Bhabha National Institute (HBNI), Banaras Hindu University (BHU), IIT Bhubaneswar, and IIT Guwahati also emerged as partners to institutions in the US like Wayne State University, Ohio State University, the University of Tennessee, and Texas A&M University, University of Tennessee, the University of Colorado, etc. Thus, institutional level shift in Indo-US collaboration is more reflective of *changing times, changing needs* and *preferences*. How these implications can shape up India's policy perspective related to Indo-US collaboration is discussed next.

*Indo-US Research Collaboration: Policy Perspective*

Indo-US collaboration is cultivating international collaboration in science and technology. The collaboration provides the opportunity to leverage scientific and technological discoveries, such as international datasets and expertise, enhance international cooperation through joint projects, and participate in multinational standards through joint partnerships. The Indo-U.S. Science and Technology Forum (IUSSTF), which was established in March 2000 as a result of an agreement between the governments of India and the United States of America, is an independent bilateral organisation that receives funding from both governments. Its mission is to advance science, technology, engineering, and innovation through meaningful dialogue between the three sectors of government, academia, and business. There are also some other bilateral cooperation programs running between the two countries. Thus, as one of the concluding remarks, we emphasize that the IUSSTF and Indian nodal agency can take steps to encourage investigation on the major factors that led to an overall trend of Indo-US collaboration share decline at individual and institutional level. We have provided a starting point for such investigation in this work. Rigorous investigations in this direction may help in formulation of new policies and associated strategies and programmes for reviving the strength of Indo-US partnership, as US is found to be a highly rewarding collaboration partner in terms of research impact. Proper emphasis should be laid to find out new areas/subject categories of mutual interest and key themes of current as well as futuristic relevance within those categories to make maximum benefit out of Indo-US collaborations.